\documentclass{elsart}
\usepackage{epsfig}
\usepackage{rotate}
\usepackage{amssymb}
\usepackage{amsmath}
\usepackage{color}

\newcommand\pp{{\gamma '}}
\newcommand\g{{\gamma }}
\newcommand\gpp{{\gamma \to \pp }}
\newcommand\ppg{{\pp \to \gamma }}

\begin{document}

\flushright{DESY 08-046 \\ \hspace{1cm} \\ \hspace{1cm} }

\begin{frontmatter}
\begin{center}
{\large \bf On search for eV hidden sector photons\\
 in  Super-Kamiokande and CAST experiments. }
\end{center}
\vspace{0.5cm}

\begin{center}
Sergei N.~Gninenko$^a$\footnote{E-mail address:
 Sergei.Gninenko\char 64 cern.ch} and 
Javier Redondo$^b$\\
{\it $^a$Institute for Nuclear Research of the Russian Academy of Sciences,\\
 117312 Moscow, Russia\\
$^b$ Deutsches Elektronen-Synchrotron DESY\\
Notkestrasse 85, D-22607 Hamburg, Germany}
\end{center}

\begin{abstract}
If light hidden sector photons ($\pp$s) exist, they 
could be produced through kinetic mixing with  solar photons in the eV 
energy range. We propose to search for this hypothetical $\pp$-flux with the  
Super-Kamiokande  and/or upgraded CAST detectors. 
The proposed experiments are 
sensitive to the $\gamma - \pp$ mixing  strength  as small as  
$10^{-5} \gtrsim \chi \gtrsim 10^{-9}$ for the  $\pp$ mass region 
$10^{-4} \lesssim m_\pp \lesssim 10^{-1}$ eV and, in the case of 
non-observation, would improve limits recently 
obtained from  photon regeneration laser experiments  for this 
mass region.
\end{abstract}
\end{frontmatter}

\section{Introduction}

Several interesting extensions of the Standard Model (SM)  suggest
the existence of `hidden' sectors consisting of $SU(3)_C \times SU(2)_L \times U(1)_Y$
singlet fields. These  sectors   of
particles do not interact with the ordinary matter directly and  couple to 
it by gravity and possibly by other 
very weak forces.  If the mass scale of a hidden sector is too high, it  will be  
experimentally unobservable. However, there is  
 a class of models  with 
at least one additional  U$_\mathrm{h}(1)$ gauge factor where the corresponding hidden 
gauge boson could be light. For example,  Okun  \cite{okun} proposed 
a paraphoton model with a massive hidden photon  mixing 
with the ordinary photon resulting in various interesting phenomena.
A similar model  of photon oscillations has been considered by 
Georgi et al. \cite{georgi}.   
Holdom \cite{holdom} showed, that   by enlarging the standard model   by the addition of a second,
massless  photon one could construct
grand unified models which contain particles with an electric
charge very small compared to the electron charge \cite{holdom}.
These considerations have stimulated new theoretical works
and experimental tests  reported in \cite{foot1}-\cite{moh1}
(see also  references therein).

 In the Lagrangian describing the photon-hidden photon system the only allowed `connection' between the hidden sector and ours is  given by the kinetic mixing \cite{okun,holdom,foot1}  
\begin{equation}
 L_{int}= -\frac{1}{2}\chi F_{\mu\nu}B^{\mu\nu} 
\label{mixing}
\end{equation}
where  $F^{\mu\nu}$, $B^{\mu\nu}$ are the ordinary 
 and the  hidden photon  field strengths, respectively. 
 
In the interesting case when $B^\mu$ has a mass $m_\pp$, this kinetic mixing
can be diagonalized resulting in a non-diagonal mass term that 
mixes photons with hidden-sector photons. Hence, photons may oscillate 
into hidden photons, similarly to vacuum neutrino oscillations, with a vacuum mixing angle which is precisely $\chi$.

Note that in the new field basis the ordinary photon remains unaffected, 
while  the hidden-sector photon (  here denoted as $\pp$ ) is completely decoupled, i.e. do not interact with the ordinary matter at all \cite{okun,holdom,foot2}.

Experimental bounds on these massive hidden photons
 can be obtained  from searches for an electromagnetic fifth force, 
\cite{okun,c1,c2}, from   
stellar cooling considerations \cite{seva1,seva2}, and from 
experiments using the method of  photon 
regeneration \cite{phreg}-\cite{pvlas}.
Recently, new constrains on   the mixing $\chi$   for the mass region 
$10^{-4}$ eV$<m_\pp<10^{-1}$ eV have been
 obtained \cite{ring7} from the results of the
BMV \cite{bmv} and GammeV \cite{gv} collaborations. 
The new results are a factor of two better than those obtained from the previous 
BFRT experiment \cite{bfrt}.    
The Sun energy loss argument has also 
been  recently reconsidered \cite{jr1}. It has been pointed out
 that helioscopes searching for solar axions   are sensitive to the keV part of the solar
spectrum of hidden photons and the latest CAST results \cite{cast1,cast2}
have been translated into limits on the $\g - \pp$ mixing parameter \cite{jr1,jr2}.   
Strong bounds on models with additional new particles plus a $\pp$ at a low energy scale 
 could be obtained  from astrophysical 
considerations \cite{blin}-\cite{dub}.
 However, such astrophysical constraints can be relaxed or evaded in 
some models, see e.g. \cite{masso}.
Hence, it is important to perform independent  tests on the 
existence of such particles in new laboratory experiments such, for example as
ALPS \cite{alps}, LIPSS \cite{lipss}, OSQAR \cite{osqar}, and PVLAS LSW 
\cite{lsw}.\\

Since $\pp$s can be produced through mixing with real photons, 
it is natural to consider the Sun as a source of low energy $\pp$s. 
It is well known that the total emission rate of the Sun is of the order of $3.8\times 10^{26}$ W. 
Moreover, the emission spectrum is also well understood. It has a broad distribution over energies up to 10 keV,  and corresponds 
roughly to the black-body radiation at the temperature $T_0\simeq 5800$ K ($0.5$ eV).
The maximum in the solar  power   spectrum is at about   $2.5$ eV   ($500$ nm), in the blue-green part of the visible region. 
  As happens with solar neutrinos, the coherence length of the photon-hidden photon oscillations is much smaller than the distance from the Sun to the Earth.  In this case the photon-hidden photon transition probability is just $2\chi^2$ and the spectral flux of hidden photons from the solar surface will be 
\begin{equation}
\frac{d \Phi^s}{d\omega}\simeq \chi^2\  4.2\times 10^{18}\frac{\omega^2}{e^{\omega/T_0}-1}\ \frac{1}{\mathrm{eV}^3\ \mathrm{cm}^2\ \mathrm{s}}
\label{flux_surface} \ . 
\end{equation}
with the maximum at $\omega \simeq 1$ eV. 

A considerably higher contribution to the flux of hidden photons is expected from $\gpp$ oscillations in the solar interior. 
Here the usual suppression of the mixing angle due to refractive effects is balanced by a higher emitting volume and a higher temperature.
Since the suppression of the mixing is more drastic as the density increases (and maximum at the solar center) it is certainly  advantageous to
search for low mass hidden sector photons with energies
in the eV range, where the photon flux is maximal \cite{jr1}. 
For $m_\pp$ well below the eV, one can use the following conservative estimate 
for the `bulk' component of the hidden photon spectral flux at the Earth \cite{jr2}:   
\begin{equation}
\frac{d \Phi^b}{d\omega}\sim \chi^2 \Bigl(\frac{m_{\gamma '}}{\mathrm{eV}}\Bigr)^4 10^{32}\ \frac{1}{\mathrm{eV}\ \mathrm{cm}^2\ \mathrm{s}}
\hspace{1cm}  \mathrm{for} \hspace{1cm} \omega \in 1 \div 5\ \mathrm{eV} , 
\label{flux_bulk}
\end{equation}
which exceeds the surface contribution except for masses $m_\pp\lesssim 10^{-4}$ eV. 
A more detailed calculation will be given elsewhere \cite{jr3}.

In this note,  we propose  direct experimental searches for the flux of solar hidden photons.
The experiment
 could be performed with  the Super-Kamiokande neutrino detector, and/or 
in the CAST experiment at CERN upgraded with a new helioscope,  
and is based on the  photon-regeneration method 
used at low  energies.

%%%%%%%%%%%%%%%%%%%%%%%%%%%%%%%%%%%%%%%%%%%%%%%%%%%%%%%%%%%%
\section{Search for hidden photons in Super-Kamiokande}

Among detectors suitable for such kind of search the most promising one is  
Super-Kamiokande (SK) \cite{sk}. This is a large, underground, water Cherenkov 
detector located in a mine in the Japanese Alps \cite{sk}.  
The inner SK detector is a  tank, 40 m tall by 40 m in diameter.
 It is filled with 5$\times 10^4$ m$^3$ of ultra-pure water, 
the optical attenuation length $L_{abs}\gtrsim$ 70 m, and 
is viewed by 11146 photomultiplier tubes (PMT) with 
7650 PMTs mounted on a barrel (side walls) and 3496 PMTs on the top
and bottom  endcaps.

\begin{figure}
\begin{center}
  \includegraphics[width=120mm]{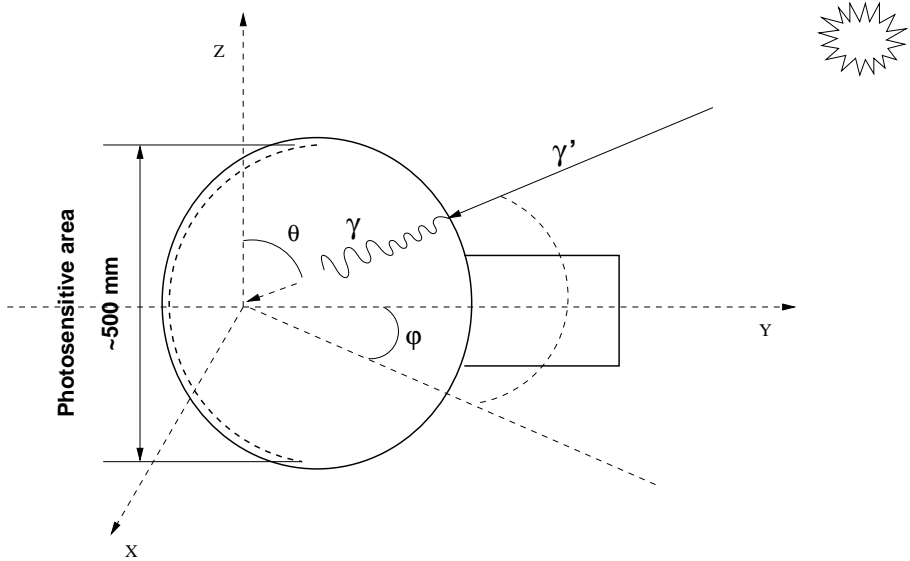}
\end{center}
 \caption{\em Schematic illustration of the direct search for light 
hidden-sector photons in the Super-K experiment.
Hidden photons penetrate the Earth and convert into visible photons
inside the vacuum volume of the Super-K PMTs. 
This results in an increase of 
the counting rate of those Super-K PMTs that are 
`illuminated' by the Sun from the back, 
in comparison with those facing the Sun. If, for instance, the Earth rotates around the Z-axis, 
the counting rate is  a periodic function of the angle $\psi$, i.e. is daily  
modulated.} 
\label{superk_PM}
\end{figure}

The PMTs (HAMAMATSU R3600-2) have  $\simeq$ 50 cm 
in diameter \cite{sk}. The full effective PMT photocathode coverage of the 
inner detector surface is 40\%.
The  photocathode, the dynode system and 
the anode are located inside a glass envelope serves as a pressure boundary 
to sustain high vacuum conditions inside the almost spherical shape PMT.  
The photocathode is made of bialkali (Sb-K-Cs) that 
matches the wave length of Cherenkov light. The quantum efficiency 
is $\simeq$ 22\% at the 
typical wave length of Cherenkov light $\simeq$ 390 nm. 
For the search for $\g - \pp$ oscillations it is important
to have the ability to see a single photoelectron (p.e.) peak,
 because the number of photons arriving at a PMT  will be   exactly one.  
 The single p.e. peak is indeed clearly seen (see e.g., Figure 9 in 
Ref. \cite{sk}) allowing to operate PMTs
in the SK experiment at a low threshold equivalent to 0.25 p.e.. It is also 
important, that 
the average PMT dark noise rate  at this threshold is just about 3 kHz.

  Since   $\pp$s are long-lived noninteracting particles, they  would   
 penetrate the Earth shielding   
and oscillate into real photons in the free space 
between the PMT envelope  and the  photocathode, as shown in 
Figure \ref{superk_PM}. The photon then would convert in the
photocathode into a single photoelectron which would be detected by the PMT.  
  This could not happen for hidden photons coming from the water tank since, as we will see, the presence of the medium suppresses $\gpp$ oscillations.  
Thus, the effect of $\pp \rightarrow \gamma$ oscillations
could be searched for in the SK experiment through 
 an increase of the counting rate of 
those PMTs that are `illuminated'  by the Sun \emph{from the back}, as shown in 
Figure \ref{superk_PM},  
in comparison with 
those facing the Sun. The increase of the counting rate in a particular
PMT depends on its orientation with respect to the Sun and is daily 
modulated. Therefore, the overall counting rate of events  from  
$\pp \rightarrow \gamma$ oscillations could also be daily modulated 
depending on the local 
SK position with respect to the Sun and  the Earth rotation axis.
 
The number $\Delta n_\g$ of expected signal events from $\ppg$ conversion  in SK  is given by an integral over time, energy band, and the surfaces of all PMTs as 
\begin{equation}
\Delta n_\g = \sum_1^N \int\hspace{-1mm} dt \int \hspace{-1mm}d\omega \ \frac{d \Phi}{d\omega}\ \eta(\omega)
\int_\mathrm{cat.}\hspace{-3mm} \vec ds \cdot {\hat r}_\mathrm{Sun} P_{\pp \rightarrow \g}(\omega)\ . 
\label{s1}
\end{equation}
Here, N$_{PMT}$ is the number of SK PMTs, $\Phi=\Phi^s+\Phi^b$ is the total
$\pp$-flux,  $\eta$ is the detection efficiency, $\vec ds$ is the photocathode surface element, ${\hat r}_\mathrm{Sun}$ a unit vector pointing to the Sun and  $P_{\pp \rightarrow \g}(\omega)$ is the $\pp \to \g$ vacuum transition probability given by: 
\begin{equation}
P_{\pp \to \g}(\omega) = 4\chi^2 \sin^2\Bigl(\frac{\Delta q l}{2}\Bigr)  
\label{prob2}
\end{equation}
where $l$ is the distance between the $\pp$ entry point to the PMT and the PMT photocathode and $\Delta q$ is
the momentum  difference between the photon and hidden photon:
\begin{equation}
\Delta q = \omega - \sqrt{\omega^2 - m_{\gamma '}^2} \approx \frac{m_{\gamma '}^2}{2\omega}
\end{equation}
assuming $m_{\gamma '}\ll \omega$. 
In the absence of photon absorption,
the  maximum of the $\ppg$ transition probability
at a distance $l$  corresponds to the case when $|\Delta q l | = \pi$. 
When $|\Delta q l| \ll \pi$ the photon and the hidden photon fields 
remain in phase and propagate coherently over the length $l$. In this 
case the transition probability degrades proportionally to $m_\pp^4$.
For example, for $\omega \simeq 3$ eV and for the maximum distance $l\simeq 50$ cm, this will occur for $m_\pp \lesssim 10^{-3}$ eV. 

The significance $S$ of the $\pp$  discovery  with the Super-K detector 
scales as  \cite{bk}
\begin{equation}
S=2(\sqrt{\Delta n_\g + n_b}-\sqrt{n_b}) 
\end{equation}
where $n_b$ is the number of expected background events.
The excess of $\ppg$ events in the Super-K detector can be 
calculated from the result of a numerical integration of eq.(\ref{s1})
over photon trajectories pointing to the PMT. 
In these calculations we use a simple model of  PMTs, 
 shown in Figure \ref{superk_PM}, 
without taking into 
account the PMT internal structure and  dead materials which might results
 in some reduction of the signal due to the photon absorption and 
damping of $\pp -\g$ oscillations.  
We also assume that the Sun is located in the plane $\Theta = \pi/2$ and 
the Earth rotates around the $Z$-axis, which is the local vertical in SK, 
  see Figure \ref{superk_PM}.
In the PMT vacuum volume
 not all $\pp$ energies effectively contribute 
to the signal because of its sinus dependence on $\Delta q$ and $l$, see 
eq.(\ref{prob2}). 
 Assuming 
the main background source is the PMT dark noise gives
\begin{equation}   
 n_b = n_0 N' t
\end{equation}
Here N' is the number of SK  PMTs contributing to the signal, and
$n_0 \simeq 3$ kHz is the average background counting rate of the PMTs 
\cite{sk}. 
Finally, taking $S=3$, $ N' \simeq 7 \times 10^3$, and  $ t\simeq 10^7$ s ($n_b=2\times 10^{14}$) results in a signal-to-background requirement of $\Delta n_\g/n_b\simeq \ 10^{-7}$.

\begin{figure}[htb!]
\begin{center}
  \includegraphics[width=100mm]{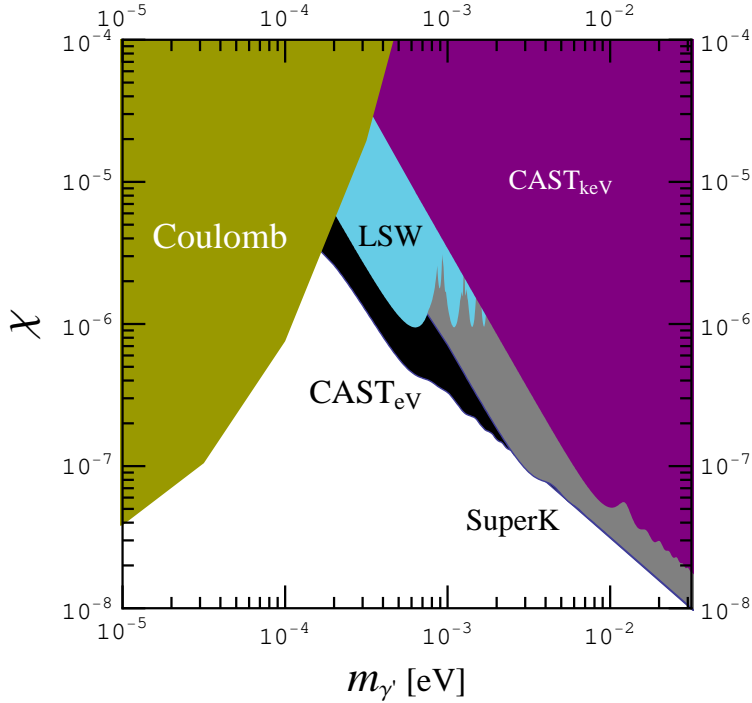}
\end{center}
 \caption{\em Regions in the ($m_\pp, \chi$) plane  which could be excluded by the proposed experiments: 
SuperK (gray region) and CAST$_{eV}$ (black). Also shown are the regions, 
with self explanatory labels,
excluded by CAST in the keV range \cite{jr1}, by LSW experiments \cite{ring7}  and by searches of deviations of Coulomb's law  \cite{c1,c2}.}
\label{plot}
\end{figure}

For the case of non-observation, 
we have computed the corresponding exclusion region  in the ($m_\pp, \chi$) 
plane shown in Figure \ref{plot}. The bound relaxes towards smaller hidden photon masses mainly because the flux \eqref{flux_bulk} is suppressed. Below $m_\pp\sim 3$ meV an additional suppression adds up because hidden photons do not have enough space to oscillate inside the SK PMT, as noted above.
The sensitive region of this experiment  surpasses the already established CAST bounds at keV energies \cite{jr1}, and for masses above $m_\pp \simeq 10^{-3}$ eV also the limits recently obtained by  Ahlers et al. \cite{ring7} from laser experiments.  

The sensitive search of $\pp$s in the SK experiment is   
possible due to unique combination of several factors, namely, 
i) the presence of the large number of PMTs with
a relatively large free vacuum volume; ii) the high efficiency of the single 
photon detection; and iii) the relatively low  PMT dark noise.
 The statistical limit on the sensitivity of the proposed experiment 
is set by the number of PMTs and by
the value of the dark noise ($n_0$) in the SK detector. 
The systematic errors are not 
included in the above  estimate, however they could be reduced by the precise 
monitoring of the PMTs gain \cite{sk}. These estimates may be strengthened  by more accurate and  detailed Monte Carlo
simulations of the proposed experiment.

Let us address the matter effects in $\gpp$ oscillations. 
Neglecting photon absorption, the $\gpp$ oscillation probability gets modified only by the refractive properties of the medium. We can parametrize them as a `photon effective mass' $m_\g$, which accounts from deviations of the photon dispersion relation, namely $\omega^2-k^2\equiv m_\g^2$, with $k$ the photon wavenumber. Consequently, we find that the mixing angle and the oscillation frequency of eq. \eqref{prob2} both get modified according to \cite{seva2} 
\begin{equation}
4\chi^2\rightarrow \frac{4\chi^2 m^4_\pp}{\left(m_\pp^2-m_\g^2\right)^2+4 \chi^2 m_\pp^4} \hspace{.3cm};\hspace{.3cm}
\Delta p\simeq \frac{m_\pp^2}{4\omega}\rightarrow \frac{\sqrt{(m_\pp^2-m_\g^2)^2+4\chi^2 m_\pp^2}}{4\omega} \ . 
\end{equation}
Therefore (as it happens with neutrinos) a high refractivity ($m_\g\gg m_\pp$) decreases the amplitude of the oscillations, a resonant conversion is possible if eventually $m_\g=m_\pp$, and the rigorous definition of the vacuum case is $m_\g\ll m_\pp$.

In a completely ionized medium like the solar interior plasma, the plasma frequency $\omega_\mathrm{P}$ plays the role of the photon mass. Here $\omega_\mathrm{P}^2=4\pi\alpha N_e/m_e$, with $\alpha$ the fine structure constant and $N_e,m_e$ the electron density and mass. Under the presence of bounded electrons the situation changes drastically, at least for photon energies around or below the atomic resonances (those relevant for us). An index of refraction $n$ is more suitable to account for the refractive properties of these media, which gives $m_\g^2\rightarrow -2\omega^2(n-1)$. The index of refraction is rarely smaller than one\footnote{The region of anomalous dispersion around the energy of an atomic transition could in principle exhibit such an unusual behavior, unfortunately light dispersion is very intense.}, so the mixing angle will be \emph{always suppressed} with respect to the vacuum case (impeding unfortunately a  resonant detection). 
	
The high vacuum conditions of the SK photo-multipliers ($p\lesssim 10^{-7}$ Torr.) make this suppression harmless, at least for $m_\pp\gtrsim 10^{-5}$ eV (using $\omega= 10$ eV and water vapor as the main residual gas, with $n-1= 2.5\ 10^{-3}$ at normal conditions). 
On the other hand, in the SK water tank, with $n\sim 1.3$ for visible light under normal conditions, the $\gpp$ oscillations get a suppression of the order $\sim m_\pp^4/(2.6\omega^2)^2$. In an optimistic case $\omega = 1$ eV and $m_\pp\sim 0.1$ eV this factor is $\lesssim10^{-4}$! Even in the  case when the SK tank is filled with air $|m_\g| \gtrsim 10^{-2}$ eV  and the sensitivity 
of the experiment could not be improved and extended to the smaller $\pp$ mass region  by searching $\g - \pp$ oscillations in the inner SK detector volume.

\section{Search for hidden photons in the upgraded CAST experiment}

The CAST (Cern Axion Solar Telescope) experiment aims to detect solar axions 
through their conversion into detectable photons in the magnetic field of 
 a 10 m long decommissioned LHC dipole magnet which is tracking the Sun. 
A detailed description of the experimental setup can be found in \cite{cast3}. 
  
\begin{figure}
\begin{center}
  \includegraphics[width=120mm]{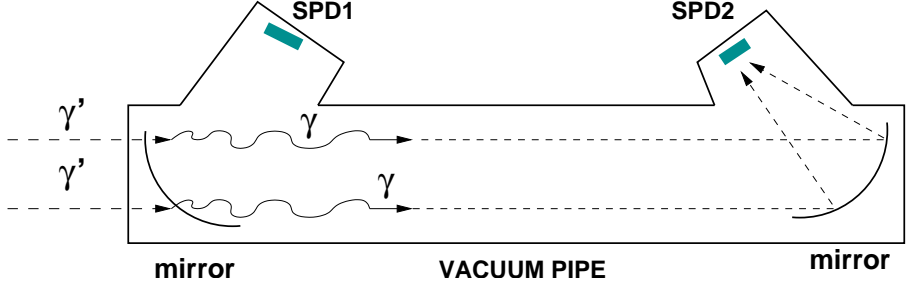}
\end{center}
 \caption{\em Schematic illustration of the direct search for solar
$\pp$-flux in the CAST experiment.  A vacuum pipe equipped  from 
both sides by 
mirrors used to focus ordinary photons produced from $\pp \to \gamma$ 
oscillations  on single photon detectors (SPDs). 
The manifestation of a signal would be  an increase of 
the counting rate of the SPD that is 
`illuminated' by the Sun 
compare to the other.} 
\label{cast}
\end{figure}

To search for eV hidden photons one can
upgrade the CAST experiment with a simple helioscope detector 
schematically shown in 
Figure \ref{cast}. Single photon detectors (SPDs), at both ends of a
vacuum pipe,  are looking for the single visible photons
 produced through oscillations of hidden photons  inside the helioscope
 when it is pointing the Sun.
 %
% The incoming hidden photon is  
%transformed into a  real, detectable photons that carries the energy and the 
%momentum of the original hidden one in vacuum pipe viewed similar to the 
%original CAST detector from the both sides by low noise single photon 
%detectors (SPD).
%
We assume that the  vacuum pipe has aperture with an effective diameter 
 of $\simeq 50$ cm and the length of 10 m. 
The whole helioscope detector  could be  mounted parallel to the LHC magnet 
on the CAST platform, allowing a movement 
of $\pm 8^o$ vertically and $\pm 40^o$ horizontally \cite{cast3}. 
This allows tracking of the Sun during about 1.5 h at sunrise and the same 
time at sunset. 
The manifestation of a signal would be an excess of events in the eV energy 
spectrum during the Sun tracking, compared to the background runs spectrum.

For a  high sensitivity of the proposed experiment to achieve a  low 
background counting rate and a high efficiency for the single photon 
detection is crucial. The SPD used     
 should have as large as possible sensitive area ($\gtrsim 1$ cm$^2$),   the 
high  red extended quantum efficiency ($\gtrsim 30 \%$),   as small as possible  noise counting rate ($\lesssim 10$ Hz), 
a high  gain ($\gtrsim 10^4$) and a good single photoelectron peak resolution. 
It is known, that one of main sources of noise, which also limits the SPD's 
single photon resolution, is 
the dark current originating from charge carriers thermally created in the 
sensitive volume of the SPD. This background  can be reduced  by the
operating the SPD at a low temperature.
Typically, the SPD dark rate decreases by  several orders 
of magnitudes when the temperature decreases from the room one  to $\simeq$ 
100 K. Thus, to achieve  a low background level ( $\simeq 1$ Hz) the SPD  
has to be cooled down to, presumably cryogenic  temperatures. Among several
 types of SPDs able to operate at such conditions,  
PMTs \cite{icarus}, silicon PMTs \cite{dolg}, large area avalanche photodiods \cite{musa}, and  a new type of hybrid photodetectors (HPD) \cite{hpd} could be  considered. 
{ In addition, optical mirrors, used to focus ordinary photons to a small 
spot, favour the use of a small size SPD detector and some
improvement of the expected signal to background ratio,   specially for the largest masses \cite{jr1}}. 
 The technique of phase shift plates proposed for axion-like particle searches \cite{Jaeckel:2007gk} could be also considered for increasing the sensitivity at higher masses.

Finally, performing integration over the spectra of eqs. \eqref{flux_surface} and \eqref{flux_bulk} results in the hypothetical CAST$_\mathrm{eV}$ exclusion region shown in Figure \ref{plot}. 
The calculations are performed for the HPD quantum efficiency taken from ref. \cite{hpd} and assuming the spectral reflectivity of mirrors to be $\gtrsim 90\%$  for  the considered wavelength region.  The background counting rate is taken to be $n_0 \simeq $ 1 Hz and the exposition time $t\simeq 10^6 $ s. Note that
the signal-to-background requirement of   $\Delta n_\g/n_b \simeq  10^{-3}$ is 
 significantly lower as compared to the  SK case.

One can see that the proposed experiment improves the sensitivity of the SuperK at low masses due to the extra oscillation length. Around $m_\pp =0.2$ meV, the hidden photon flux from the solar interior is strongly suppressed and the sensitivity of the experiment is dominanted by the surface contribution of eq.\eqref{flux_surface}.
Moreover, it surpasses the reach of already performed laser experiments \cite{ring3} and certainly of the CAST keV search \cite{jr1}. The vacuum requirements are somehow crucial. Using again water vapor as a possible residual gas 
in the helioscope pipe, we 
should ensure a pressure below $\simeq 10^{-4}$ Torr not to dump oscillations of $\gtrsim$ eV hidden photons with $m_\pp>2\ 10^{-4}$ eV.

%This proposed experiment could run for the second phase of the CAST experiment
% (2008-2010) exploring the parameter space in the 
% ($m_\pp, \chi$)  plane  shown in Figure \ref{plot}.
 
{\large \bf Acknowledgments}

We would like to  thank K. Zioutas and the CAST collaboration 
 for their interest to this work and useful discussions.
S.N.G. is grateful to  N.V. Krasnikov, V.A. Matveev and V. Popov for useful
comments,  A. Korneev for help in calculations and Yu. Musienko for discussions
on low noise photodetectors.  
J.R. would like to thank the people of the ALPS collaboration for fruitful discussions, the German SFB 676 project C1 for funding and specially to S. Gninenko for the invitation to join this study and a nice communication during 
the project. This work was supported by Grants RFFI 07-02-256a and 
RFFI 08-02-91007-CERNa.

\end{document}